\documentstyle[preprint,aps]{revtex}
\begin{document}
\draft
\title{Magneto-Conductance Anisotropy and\\
Interference Effects in Variable Range Hopping}
\author{Ernesto Medina}
\address{Intevep SA, Apartado 76343,
Caracas 1070A, Venezuela}
\author{Mehran Kardar}
\address{Department of Physics, Massachusetts Institute of
Technology, Cambridge, MA 02139, U.S.A.}
\author{Rafael Rangel}
\address{Departamento de F\'\i sica, Universidad Sim\'on Bol\'\i var,
Apartado 89000, Caracas 1080A, Venezuela}
\date{\today}
\maketitle
\begin{abstract}
We investigate the magneto-conductance (MC) anisotropy in the variable
range hopping regime, caused by quantum interference effects in three
dimensions. When no spin-orbit scattering is included, there is an
increase in the localization length (as in two dimensions), producing
a large positive MC.  By contrast, with spin-orbit scattering present,
there is no change in the localization length, and only a small
increase in the overall tunneling amplitude.  The numerical data for
small magnetic fields $B$, and hopping lengths $t$, can be collapsed
by using scaling variables $B_\perp t^{3/2}$, and $B_\parallel t$ in
the perpendicular and parallel field orientations respectively.  This
is in agreement with the flux through a `cigar'--shaped region with a
diffusive transverse dimension proportional to $\sqrt{t}$. If a single
hop dominates the conductivity of the sample, this leads to a
characteristic orientational `finger print' for the MC anisotropy.
However, we estimate that many hops contribute to conductivity of
typical samples, and thus averaging over critical hop orientations
renders the bulk sample isotropic, as seen experimentally. Anisotropy
appears for thin films, when the length of the hop is comparable to
the thickness.  The hops are then restricted to align with the sample
plane, leading to different MC behaviors parallel and perpendicular to
it, even after averaging over many hops. We predict the variations of
such anisotropy with both the hop size and the magnetic field
strength. An orientational bias produced by strong electric fields
will also lead to MC anisotropy.
\end{abstract}
\narrowtext
\section{Introduction}
\label{secintro}
Striking quantum interference (QI) effects have been observed in
experiments on {\it insulating}
materials\cite{rOrlov,rOvadyahu,rMilliken,rSarachik}.  These
observations are of particular interest, as they point to quantum
coherence phenomena for strongly localized electrons, where naively
they may not have been expected to occur over length scales
appreciably larger than the localization length $\xi$. A theoretical
explanation was first proposed by Nguyen, Spivak and Shklovskii
(NSS)\cite{rNSS} in the context of Mott Variable Range Hopping
(VRH)\cite{rMott}: Phase coherence is maintained over the long
distances between phonon assisted tunneling events, which grow with
decreasing temperature $T$ as $\exp(T_0/T)^{1/(D+1)}$ in $D$ spatial
dimensions. The resulting coherence length can be quite large
(typically of the order of $20-50\xi$).  In this work, we consider the
three dimensional NSS model and focus on the dependences of the
conductance and its fluctuations on the relative orientations of the
magnetic field and the dominant hop.

The initial indications of QI came from observations of a strong
positive magneto-conductance (MC) in materials that exhibit VRH
behavior\cite{rbunchII}.  In a single impurity picture, the action of
the magnetic field is to further confine electrons already localized
around the impurity.  This would result in a negative MC which is not
the case in experiments for weak magnetic fields. Further evidence is
provided by the orbital nature of the MC\cite{rOvadyahu} observed in
$InO$ films of varying thickness. While experiments show an isotropic
MC for thick samples, anisotropy sets in when the film thickness is
close to the Mott hopping length. Such anisotropy precludes
explanations in terms of scattering of electron spin by magnetic
impurities\cite{rJapones}, which are necessarily isotropic with
respect to the field direction, pointing instead to interference
effects due to the electron orbits.  Finally, in a careful set of
experiments, Orlov et al\cite{rOrlov} and Milliken and
Ovadyahu\cite{rMilliken} demonstrate the presence of reproducible
conductance fluctuations or magneto-fingerprints, generally regarded
as a clear signature of QI effects.

The NSS model considers the QI between the many virtual paths that the
electron can take while tunneling under the barrier between two
distant impurity centers. In the tunneling process, the hopping
electrons with energies near the Fermi level, encounter impurities
with energies outside the Mott energy strip. These impurities are
considered as the source of elastic scattering events under the
barrier. Since the contribution of each virtual path (tunneling
through a barrier) is exponentially damped by the distance it covers,
it is sufficient to ignore back-scattering and focus on the (directed)
paths that only undergo forward scattering between the initial and
final impurities.

The initial numerical studies of the NSS model (on relatively small
systems)\cite{rNSS} indeed confirmed that it yields the correct sign
for the MC.  Subsequently, Sivan, Entin-Wohlman, and Imry\cite{rSEI}
provided a theoretical analysis that agrees with much of the early NSS
results.  The critical hop is identified from the condition of
producing a percolating network of random resistors\cite{rAHL}, while
the probability distribution for individual hops is calculated by
assuming that the contributing virtual paths are uncorrelated. The
latter assumption, which we shall refer to as the Independent Path
Approximation (IPA), was shown to be invalid by Shapir and
Wang\cite{rShWa}, since, in low dimensions, the paths must intersect
at some scattering sites.  Eventually, the correct form of the hopping
probability distribution was calculated by Medina et
al\cite{rMedinaNoSO}, by incorporating the correlations between the
virtual paths.  The analytical results, confirmed by extensive
numerical simulations on very large sizes, indicate that the positive
MC in this model actually corresponds to an increase in the
localization length with the magnetic field in the absence of
Spin-Orbit scattering. This prediction is supported, at least
qualitatively, by recent experiments on $InO$\cite{rOvadyahu} and
$YSi$\cite{rSanquer}.  While the IPA scheme cannot produce a change in
the localization length, an alternative approach to strong
localization, based on Random Matrix Theory (RMT)\cite{rSanquer}, also
produces such an effect.  The latter, which is exact only for
quasi-one dimensional systems, predicts a doubling of the localization
length.

There are conflicting theoretical and experimental observations in the
presence of Spin-Orbit scattering. The first experimental study on
$InO$\cite{rShapOvad} showed MC behavior resembling that of the weak
localization regime; i.e. a positive MC for low fields, changing to
negative at higher fields. On the other hand, more recent
experiments\cite{rSanquer} on $YSi$ show negative MC for all applied
fields. On the theoretical side, both the IPA scheme\cite{rYigal}, and
the correct accounting of correlations\cite{rMedinaSO}, yield a
positive MC without changes in the localization length.  (In fact the
IPA results for MC are exact in the presence of strong SO.)  Finally,
the Random Matrix\cite{rSanquer} approach finds a negative MC caused
by a universal decrease of the localization length by a factor of two.
These issues are discussed in greater detail in
reference\cite{rMedinaRev}, and will not be discussed further in the
present work.

In this work we study the NSS model in three dimensions, with and
without spin-orbit scattering. An important new feature is that we
must now take into account the relative directions of the magnetic
field and the dominant hop. This issue is most relevant experimentally
for samples that are small enough (or at such low temperatures) to
include only a single dominant hop. By measuring the MC anisotropy as
a function of the direction of the magnetic field, it is possible to
locate the orientation of this dominant hop! Field dependences
parallel and perpendicular to this orientation can then be used to
further test the current models of coherence in the localized regime.
There is, however, a certain amount of internal averaging when the
conductance is dominated by several hops.  Some insight about the
nature of the hops can then be obtained by examining the MC anisotropy
of thin films, as a function of their thickness and orientation to the
magnetic field.

\section{The NSS Model}
\label{secNSS}
Low temperature conduction in the strongly localized regime is
dominated by thermal hopping. At the lowest temperatures localized
electrons lack enough thermal energy to hop to neighboring sites. On
the other hand, electrons cannot wander too far away from their
localization point due to the exponential decay of the wave function.
The balance of these competing tendencies results in an optimal
hopping length and leads to Mott's law for VRH\cite{rMott}.  Each of
these hops may be represented by an effective resistor (hopping
probability) in a network which can then be solved for the macroscopic
conductance of the sample. The picture of the Miller-Abrahams
(MA)\cite{rMillerAbra} network is central to the understanding of
hopping conduction.  The effective resistance of the MA network was
first estimated by Ambegaokar, Halperin, and Langer (AHL)\cite{rAHL},
and Shklovskii and Efros\cite{rSKEFpases}, using a percolation
argument: Due to the exponentially large values of the resistors in
the network, the macroscopic conductance is dominated by a single
bottle-neck resistor on a percolating network. (We shall later discuss
the modifications due to multiple hops).  This simple argument
provides a powerful tool, since it is then sufficient to determine the
variations of a single hop with various external (applied fields) or
internal (doping, correlation effects, anisotropy) physical
parameters\cite{rEfrosShk}.

The model proposed by NSS examines QI effects for the dominating hop.
Due to the long distance of the hop, typically $R\sim
\xi(T_0/T)^{1/(D+1)}\sim 20-50\xi$, electrons scatter off many
impurities on route to the final site.  While at the end of the
process there is some loss of phase coherence (due to inelastic
scattering by phonons), the intermediate scattering is elastic. To
study QI processes for the hop, the NSS model places the impurities on
the sites of a regular lattice; e.g. the cubic lattice in Fig.\
\ref{fig1}. The interference effects are maximized if the initial and
final sites for the hop are chosen at diagonally opposite end-points.
Electrons can then follow many different virtual paths from the
initial to the final site.  The overall tunneling amplitude is
computed by summing all (virtual) paths between the two points,
each contributing an appropriate quantum mechanical complex weight.
These weights are obtained from an Anderson tight-binding Hamiltonian
\begin{equation}\label{eATH}
{\cal H} = \sum _{i} \epsilon _{i} a^{+} _{i} a _{i} +
\sum _{<ij>} V_{ij} a^{+} _{i} a _{j},
\end{equation}
where $\epsilon_{i}$ are the impurity site energies, and $V_{ij}$
represent the nearest neighbor couplings or transfer terms. NSS
further simplify the problem by choosing site energies distributed
according to $$\epsilon_i= \cases{+W &with probability $p$,\cr -W
&with probability $(1-p)$,\cr}$$ and a transfer term $$V_{ij}=
\cases{V &if $i,j$ are nearest neighbors,\cr 0 &otherwise.\cr}$$ We
shall henceforth set $p=1/2$. To describe strong localization, the
Anderson parameter is taken to be much smaller than one ($V/W \ll1$).
This corresponds physically to a strongly disordered sample where the
width of the bands ($\sim 2V$) centered at energies $\pm W$ is much
smaller than their energy difference to the Fermi level.

The effective hopping matrix element can be computed using a locator
expansion\cite{rAnderson}\cite{rMedinaRev}. The overlap amplitude
(Green's function) between the initial and final sites is given by
\begin{equation}\label{eAGFa}
\langle\Phi|G(E)|\Psi^+\rangle
= \sum_{\Gamma}\prod_{i_{\Gamma}}{Ve^{iA}\over E_f-
\epsilon_{i_{\Gamma}}},
\end{equation}
where $|\Phi\rangle$ represents the state with a localized electron at
the initial site, and $|\Psi^+\rangle$ the state with a localized
electron at the final site; $E_f$ is the Fermi energy which will be
set to zero, and $A$ is the magnetic vector potential.  In principle,
the sum is over {\it all} paths $\Gamma$ between the initial and final
sites (including back-scattering).  However, for $V/W \ll1$, only the
shortest (forward scattering) paths need to be included. (For a more
detailed discussion and justification on this point, using an analogy
with high temperature expansions in the Ising model, see
\cite{rMedinaRev,rMKlh}.) Neglecting back-scattering, we obtain for
paths of length $t$,
\begin{equation}\label{eAGFb}
\langle i|G(E)|f\rangle =\left({V \over W}\right)^{t}  J(t),
{}~~J(t)=\sum^{\scriptstyle directed}_{\Gamma'}\prod_{i_{\Gamma'}}
\eta_{i_{\Gamma'}}e^{iA}.
\end{equation}
The sum is now restricted to directed paths $\Gamma'$, and
$\eta_i={\rm sign}\left(\epsilon_i\right)=\pm 1$. The interference
information is captured in the function $J(t)$, while the factor
$(V/W)^t$ is the leading contribution to the expected exponential
decay of the localized wavefunction.

The transfer matrix approach provides an efficient numerical algorithm
for computing $J(t)$. As described in reference\cite{rMKlh}, this
method allows summing over the exponentially large number of paths in
polynomial time (typically $\sim t^D$ for $D$ dimensions).  The
results of extensive analytical and numerical studies (mostly in
$D=2$) based on this method are discussed in
reference\cite{rMedinaRev}.  Briefly, the probability distribution for
$J(t)$ is quite broad. Its {\it logarithm} is a universal function
with a mean is proportional to $t$, and variance growing as
$t^{2\omega}$, with $\omega$ depending on the dimension $D$. Since the
mean and variance of the (log-) distribution are independent, {\it two
parameters} are necessary to describe the tunneling probability. High
moments of the distribution are however nonuniversal, and dominated by
exceptionally good samples\cite{rMedKarNonU}.

In discussing the change in the tunneling probability in a magnetic
field $B$, NSS introduced\cite{rNSS} the important concept of the
effective `cigar'--shaped area through which the field penetrates.
Naively, typical directed paths execute a random walk in the
transverse direction, so that a path of length $t$ wanders away a
distance of the order of $t^{1/2}$. As shown in Fig.\ \ref{fig2} the
area presented to a magnetic field perpendicular to such paths thus
grows as $A_\perp\propto t\times t^{1/2}\propto t^{3/2}$, and the MC
is expected to be a function of the flux $Bt^{3/2}$. This is indeed
the case in the absence of randomness, where the exact response of the
sum over directed paths in $D=2$\cite{rMedinaNoSO} initially decreases
as $B^2t^3$. The above argument does not work in the presence of
randomness, where typical paths have super-diffusive transverse
fluctuations that grow as $t^\zeta$ with $\zeta>1/2$\cite{rMKlh}.
However, the scaling functions {\it are not} simply modified to depend
on $Bt^{1+\zeta}$. In the presence of spin--orbit (SO) scattering, the
behavior is qualitatively similar to the pure case: there is a
positive MC, initially scaling as $B^2t^3$, which saturates at a
finite ($t$ independent) value. By contrast, in the absence of SO, the
(positive) MC grows unbounded with $t$. This is because the effect of
the magnetic field is a (nonuniversal) increase in the localization
length $\xi$, initially scaling as $B^{1/2}$. The appropriate scaling
variable in this case is $Bt^2$, although numerically one finds a
small pre-asymptotic regime with $Bt^{3/2}$ scaling. There is
currently no satisfactory explanation of the crossovers in the absence
of SO.

The replica arguments\cite{rMedinaRev} suggest that the same
asymptotic behavior for the MC should be observed in $D=3$, as long as
the magnetic field is perpendicular to the hopping direction. However,
it is also possible to consider fields parallel to the hop. The
transverse area presented to the magnetic field by typical diffusive
paths (see Fig.\ \ref{fig2}) now grows as $A_\parallel\propto
t^{1/2}\times t^{1/2}$, suggesting $B_\parallel t$ as the appropriate
scaling argument. This simple argument was first presented in
ref.\onlinecite{rEKR}, along with preliminary numerical support.  The
anisotropic field dependence was verified recently by Lin and
Nori\cite{rNori} in the IPA approach.  In the next section we present
detailed numerical results pertaining to the anisotropy of MC in
$D=3$.

\section{Numerical Results for a Single Hop}
\label{secResults}
As the distribution of the tunneling amplitude $J(t)$ in
eq.(\ref{eAGFb}) is broad, care in averaging is quite important.
We typically averaged the logarithm of the transition probability
(log-conductance) over 2000 realizations of randomness. The transfer
matrix method allows us to examine systems of size $t=600$ in the
wedge geometry.  Furthermore, after studying the dependence of the
computed amplitudes on the lateral dimension, we also used a bar
geometry with dimensions $1500\times 200\times 200$. This is
reasonable if the important paths have transverse fluctuations smaller
than 200, which was found to be the case.  As discussed before, our
main focus is on the different responses for fields parallel and
perpendicular to the hop direction. We discuss separately the MC with
and without SO scattering.  The results in the presence of SO are
easily interpreted and offer no fundamental surprises. However, most
of our numerical results in the absence of SO pertain to a
pre-asymptotic regime for which we have no satisfactory theoretical
understanding, but which are most probably of experimental importance.

\subsection{MC without Spin-Orbit Scattering}

Fig.\ \ref{fig3} shows the MC and its fluctuations for a magnetic
field {\it parallel} to the hopping direction. For the largest values
of $B$, it is clear that $\ln (|J(t,B)|/|J(t,0)|)^2$ grows linearly
with the length $t$ of the hop. This is indicative of an exponential
correction to the conductance due to an increased localization length
in the magnetic field. It is only after about $t=400$ that reasonable
linearity is achieved, so rather large systems must be examined to
study the true asymptotic limit. A similar positive MC (and increased
localization length) behavior is also observed for the perpendicular
field orientation. Concurrently, there is a reduction in the magnitude
of the fluctuations in the tunneling probability (inset Fig.\
\ref{fig3}), and there appear to be strong correlations between
changes in the average of the log-conductance and its fluctuations.
This is also the case in $D=2$, where a replica argument suggests that
these two quantities (indeed the whole probability distribution) are
governed by a single parameter\cite{rMedinaRev,rMKlh}.

Most of the data in Fig.\ \ref{fig3} pertains to a pre-asymptotic
regime, before the change in localization length becomes apparent.
Since the length of the hop in most experiments is only of the order
of 20 to 50 $\xi$, it is useful to explore this regime carefully. In
Fig.\ \ref{fig4} we present an attempt to collapse the numerical data
in this regime for different values of $B$ and $t$.  The collapse for
the parallel field orientation is demonstrated in Fig.\ \ref{fig4}a;
the maximum hopping length in this graph is $t=600$, while fields go
up to $0.1$ flux quanta per plaquette; all in the pre-asymptotic
regime.  The choice of the scaling variable $B_\parallel t$, is
consistent with the flux through a section of the NSS `cigar'
perpendicular to the hop direction ($A_\parallel\propto
\sqrt{t}\times\sqrt{t}$).  Two regimes are apparent in Fig.\
\ref{fig4}a: (i) for the lowest fields ($5\times10^{-5}\phi_o$ per
plaquette) and sizes ($t=10-100$) there is a linear increase with the
variable $Bt$. (ii) For intermediate fields and hop sizes, when
approximately one flux quantum penetrates the NSS `cigar,' there is a
non-trivial apparent exponent. The behavior in these regimes is
summarized by
\begin{equation}\label{eScaPA}
\left\langle\ln {|J(t,B_{\parallel})|\over
|J(t,0)|}^2\right\rangle = \cases{1.5B_{\parallel}t &for
$B_{\parallel}t\leq 1$\cr (a_1B_{\parallel}t)^{\alpha_1} &for
$B_{\parallel}t > 1$\cr},
\end{equation}
where $\alpha_1= 0.38\pm 0.02$.

The corresponding collapse for fields in the perpendicular orientation
is presented in Fig.\ \ref{fig4}b. In this case the appropriate
scaling variable is $Bt^{3/2}$, again consistent with the flux through
the NSS `cigar.'  Once more, two different regimes are identified,
with
\begin{equation}\label{eScaPE}
\left\langle\ln {|J(t,B_{\perp})|\over
|J(t,0)|}^2\right\rangle = \cases{0.6B_{\perp}t^{3/2} &for
$B_{\perp}t^{3/2}\leq 1$\cr  (b_1B_{\perp}t^{3/2})^{\alpha_2} &for
$B_{\perp}t^{3/2}>1$\cr},
\end{equation}
and $\alpha_2= 0.25\pm 0.02$.  We again emphasize that the second
regime above is still pre-asymptotic. For larger systems ($200\times
200\times 1500$) the log-conductance crosses over to a regime where
presumably the relevant scaling variable is $Bt^2$\cite{rMedinaRev}.
The latter scaling suggests that the magnetic length is the
relevant length scale\cite{rSpivak}. We were not able to clearly
access this regime as cumbersomely large systems must be simulated.

It is interesting to compute, on the basis of the above results, the
anisotropy in conductance of a single critical resistor. We shall
define an anisotropy parameter, $$\beta(B,t)={\left\langle\ln
|J(t,B_{\perp}=B)|^2-\ln |J(t,0)|^2\right\rangle\over
\left\langle\ln |J(t,B_{\parallel}=B)|^2-\ln |J(t,0)|^2
\right\rangle},$$
thus making contact with the original experimental definition of
ref.\onlinecite{rOvadyahu}. Depending on the strength of the magnetic
field, this anisotropy shows different scaling forms: For the smallest
fields, such that $Bt^{3/2}< 1$, $$\beta={0.6Bt^{3/2}\over
1.5Bt}=0.4t^{1/2}.$$ In this range the anisotropy is field
independent, but changes with temperature since $t=\xi(T_0/T)^{1/4}$.
There is an intermediate regime where $Bt^{3/2}> 1$ while $Bt< 1$, and
$$\beta\propto B^{\alpha_2-1} t^{1.5\alpha_2-1}, $$ which depends on
both $B$ and $t$.  Finally, for $Bt>1$, $$\beta\propto
B^{\alpha_2-\alpha_1}t^{1.5\alpha_2-\alpha_1},$$ which, using the
numerically estimated values, is approximately independent of $t$, and
has a weak field dependence as $\beta\propto 1/B^{0.13}$.  Thus
anisotropy is reduced when the field increases as shown in Fig.\
\ref{fig5}.  Qualitatively similar behavior is observed in $InO$
samples for sufficiently high fields in ref.\onlinecite{rOvadold}.

\subsection{MC with Spin-Orbit Scattering}

When spin-orbit active impurities (doping with heavy elements) are
taken into account, the NSS model must be generalized to include
scattering of the spins. The tight binding
Hamiltonian is now modified to
\begin{equation}\label{eHSO}
{\cal H}=\sum_{i,\sigma}\epsilon_i
a^\dagger_{i,\sigma}a_{i,\sigma} +\sum_{<ij>,\sigma \sigma'}
V_{ij,\sigma \sigma'} a^\dagger_{i,\sigma}a_{j,\sigma'}\qquad,
\end{equation}
where $\sigma$ indicates the electron spin.  The constant
nearest-neighbor hopping elements $V$ in eq.(\ref{eATH}) are no longer
diagonal in spin space. Instead, each is multiplied by $U_{ij}$, a
randomly chosen $SU(2)$ matrix which describes the spin rotation due
to strong SO scatterers on each bond\cite{rYigal,rZP}.
Eq.(\ref{eAGFa}) for the overlap of wave-functions at the two
end-points must now include the initial and final spins, and has the
locator expansion
\begin{equation}\label{eAGFso}
{\langle i\sigma|G(E)|f\sigma'\rangle =\sum_{\Gamma}
\prod_{i_\Gamma} {V  e^{iA}U\over E_f-\epsilon_{i_\Gamma}}.}
\end{equation}
Each bond along the path contributes a random spin rotation $U$, and a
phase factor from the magnetic vector potential $A$, resulting in
\begin{eqnarray}\label{eAGso}
{\cal A}=\langle i\sigma|G(0)|f\sigma'\rangle =W(V/W)^tJ(t) ;\nonumber
\\ {\rm with}\qquad J(t)= \sum_{\Gamma'} \prod_{i_{\Gamma'}} \eta_i
e^{iA}U.
\end{eqnarray}
After averaging over the initial spin, and summing over the final
spin, the tunneling probability is
\begin{eqnarray}\label{eAT}
T={1\over2}{\rm tr}({\cal A}^\dagger {\cal A})=W^2(V/W)^{2t} I(t)
;\nonumber \\ {\rm with}\qquad I(t)={1\over2}{\rm tr}(J^\dagger J).
\end{eqnarray}

For a three dimensional lattice we studied numerically the statistical
properties of $I(t)$. Using a transfer matrix we evolve paths of
length $t=600$ in the wedge geometry, averaging over $2000$
realizations of randomness. As in the previous section, we also used
the bar geometry to compute $I(t)$ for systems of size $200\times
200\times 1500$. We checked the bar-geometry results for crossover
effects (because of the smaller lateral sizes) and confirmed that
their behavior is effectively three dimensional. The numerical results
are shown in Fig.\ \ref{fig6}, and have the same qualitative features
as in two dimensions: Unlike in the absence of SO, there is no linear
increase of $\Delta\ln I_{so}(B,t)$ with system size $t$, and the MC
saturates to a field dependent value for sufficiently long hops.  The
scale of fluctuations in (the logarithm of) $I(t)$ is not
significantly modified by the $B$ field. As in $D=2$, turning on the
SO scattering from zero, thus switching from an orthogonal to a
symplectic Hamiltonian, is accompanied by an increase in the zero
field conductance, and a concomitant reduction in conductance
fluctuations.

Fig.\ \ref{fig7} demonstrates the collapse of the MC data for fields both
parallel and perpendicular to the hopping direction.  For fields
parallel to the hopping direction the appropriate scaling parameter is
again $B_\parallel t$, corresponding to the flux through the area
$A_\parallel \propto (\sqrt{t})^2$. The scaling function has the form
\begin{equation}\label{eScaSOPA}
\langle\ln I(t,B_{\parallel})-\ln I(t,0)\rangle =
\cases{0.4B^2_{\parallel}t^2 &if $B_{\parallel} t<1$\cr C\approx 0.2
&if $B_{\parallel}t>1$\cr}.
\end{equation}
For fields in the perpendicular orientation, the appropriate area is
$A_\perp\propto t^{3/2}$ (see Fig.\ \ref{fig2}), leading to a scaling
function
\begin{equation}\label{eScaSOPE}
\langle\ln I(t,B_{\perp})-\ln I(t,0)\rangle =
\cases{0.1B^2_{\perp}t^{3} &if $B_{\perp} t^{3/2}<1$\cr C\approx 0.2
&if $B_{\perp}t^{3/2}>1$\cr}.
\end{equation}
The saturation value, when one flux quantum threads the appropriate
area, is roughly the same in the two cases.

The anisotropy parameter in the presence of SO is $$\beta=\cases{0.25t
&if $Bt^{3/2}<1$\cr 1 &if $Bt>1$\cr},$$ with a small crossover region
between $Bt^{3/2}\sim1$ and $Bt\sim1$. We thus obtain a hopping size
dependent anisotropy (which might show up as a temperature dependent
anisotropy) for low fields. For higher fields, anisotropy in the
presence of SO scattering disappears as the field is increased beyond
a flux quantum through the smaller of the typical areas found above.
In order to properly compare with experiments, one must average over
different magnetic fields orientations as discussed in the next
section.

The MC in the presence of SO can be explained by a replica analysis.
After averaging over the impurity potential $\epsilon_i=\pm W$, the
moment $\langle I(B,t)^n\rangle $, is obtained as a sum over $n$
paired paths. The pairings either involve paths taken from $J$ and
$J^\dagger$; the contributions from the magnetic vector potential
cancel for such `neutral' pairs, which do not contribute to the MC. It
is also possible to construct pairs with both paths taken from either
$J$ or $J^\dagger$; such `charged' paths are responsible for magnetic
response.  An interesting feature of averaging over strong SO
scatterers (the matrices $U$ in eq.(\ref{eAGso})) is that the charged
and neutral paths become completely decoupled and can be treated
independently\cite{rMedinaRev}.  Thus the MC with SO is (rather
fortuitously) calculated exactly by IPA: there is no change in the
localization length, and only a constant increase in the tunneling
amplitude. Recently Lin and Nori\cite{rNori} computed the field
dependencies in a scheme that is equivalent to IPA in two and three
dimensions. (The equivalence to IPA can be readily recognized from the
equality of the moments of the `conductance' to those of a Gaussian
obtained in ref.\onlinecite{rSEI}.) In the small field limit Lin and
Nori confirm exactly the scaling variables derived by us\cite{rEKR}
here, for the two field orientations.
Another notable feature of the data is the $B^2$ dependence for small
fields, as expected in the IPA approach\cite{rSEI}.

\section{Averaging over Many Hops}  \label{secAverage}

So far we focused on the response of a single hop to a magnetic field.
However, the percolation arguments\cite{rAHL}\cite{rSKEFpases} for the
Miller-Abrahams network leads to the conclusion that critical
resistors dominate only over a correlation length $\ell$. Starting
from the properties of a percolation cluster near the threshold it is
concluded\cite{rEfrosShk} that a single hop is responsible for the
overall conductivity only for length scales up to
\begin{equation}\label{eDisleng}
\ell\sim\xi\left ({T_0\over T}\right )^{{\nu+1\over D+1}},
\end{equation}
where $\nu$ is the exponent for the divergence of the correlation
length close to the percolation threshold. The macroscopic system is
then built by superposing many blocks of length $\ell$.  Therefore, in
general, many critical hops, in general, contribute to the
conductivity of a large sample. In eq.(\ref{eDisleng}) the variable
$~T_0/T$ can be regarded as a measure of disorder strength; it
increases when the density of states at the Fermi level, or the
localization length, decreases. As disorder increases, the volume
dominated by the critical NSS resistor gets larger.

For samples of size $L\gg\ell$, many hops contribute to the
conductivity, and the overall MC must be calculated from their {\it
average} response.  Using the experimentally reported\cite{rOvadyahu}
data, we can estimate $\ell$: For samples of thickness $d=100$\AA\ and
hopping length $t\approx 3-4 d$, one obtains $\ell\approx 1000$\AA.
Although this length is large compared to the sample thickness, it is
still very small in comparison to the sample planar dimensions
($10\times 6$ mm for the experiments in reference\cite{rOvadyahu}).
Many hops thus influence the conductance of these samples, each with a
presumably different orientation with respect to the magnetic field.
In the following paragraphs we shall perform an average over the
directions of the NSS `cigar'--shaped' region with respect to the
magnetic field. We shall assume that the electric field is
sufficiently weak so that the relative orientations of the hops are
still randomly distributed in
space\cite{rMillerAbra,rNguyenShklovskii}.  We expect that for thick
samples (as compared to the hopping length $t$) the averaging over all
possible directions of the hop removes the anisotropy in MC, as seen
in experiments. On the other hand, for thin samples such that $d\leq
t$, the restricted averaging over effectively two-dimensional hops
should lead to significant field anisotropy. Such thickness dependent
anisotropy is indeed observed in the experiments of Faran and
Ovadyahu\cite{rOvadyahu}, and those of Orlov and
Savchenkov\cite{rOrlov,rbunchII}.

We shall assume that the appropriate `cigar' shaped region for
calculating magnetic interference phenomena is an ellipsoid of
revolution as depicted in Fig.\ \ref{fig8}.  The major and minor axes
of the ellipsoid are denoted by $b$ and $a$ respectively. Following
NSS, $2b\equiv t$ is the length of the dominant hop, while
$a=\sqrt{t\xi}$ is a typical diffusive distance in the transverse
direction.  Consider a magnetic field $B$ at an angle $\theta_t$ with
respect to the major axis. We take the relevant magnetic flux to be
that which penetrates the projection of the ellipsoid onto a plane
perpendicular to the $B$ field, as indicated in Fig.\ \ref{fig8}. This
projection is an ellipse of minor axis $a$, and with a major axis
of length $c=\sqrt{a^2\cos^2\theta_t+b^2\sin^2\theta_t}$.  In previous
sections we demonstrated that in the weaker field regimes $\Delta\ln
|J(B,t)|^2\propto \langle|\Phi|^{\gamma}\rangle$, where $\Phi$ is the
appropriate flux. In the presence of SO, $\gamma=2$ as justified by
IPA, while $\gamma=1$ in the absence of SO from the numerical results.
In the following, we shall focus on $\gamma=2$ with SO, which has a
better justification, and for which it is easier to compute the
average,
\begin{eqnarray}\label{eflux}
\left\langle\Phi ^2\right\rangle&=&\left\langle \left( \pi ac B
\right)^2 \right\rangle\nonumber\\
&=&\pi^2 B^2 a^2\left[ b^2-\left( b^2-a^2\right)\left\langle
\cos^2\theta_t\right\rangle\right].
\end{eqnarray}

In performing the average over hop orientations $\theta_t$, we
distinguish between the following regimes:

\par\noindent{\bf (i)}
The behavior of samples of thickness $d\gg t$ is effectively three
dimensional.  All orientations $\theta_t$ are equally likely (in low
voltage bias), and
$\left\langle \cos^2\theta_t\right\rangle=1/3$. The averaged response
is isotropic and depends upon
\begin{equation}\label{ethickB}
\langle\Phi^2\rangle={B^2\pi^2 a^2\over 3}(2b^2+a^2).
\end{equation}
The characteristic $B^2$ dependence is a signature of the IPA, and is
also observed experimentally for small fields. In fact, the flux
through a typical interference region in ref.\onlinecite{rOvadyahu} is
of the order of $0.5 hc/e$ which corresponds to the regime
$Bt^{3/2}<1$.

\par\noindent{\bf (ii)} For $a\ll d \simeq t$, the range
of orientations of the cigar is limited by the finite thickness,
leading to MC anisotropy.  Consider a magnetic field at an angle
$\alpha$ with respect to the plane of the sample. We shall indicate
the orientation of the major axis of the ellipsoid by a polar angle
$\theta$ (with respect to the normal to the sample), and an azimuthal
angle $\phi$. Because of the finite thickness, the range of $\cos
\theta$ is limited to the interval $\left[-d/2b,+d/2b\right]$, and
allowed angles in this range are weighted by
\begin{equation}\label{ptheta}
p(\cos\theta)=\cases{{2b/ d}-{4b^2}|\cos\theta|/d^2 &if
$|\cos\theta|<d/2b$\cr 0&if $|\cos\theta|>d/2b$\cr}.
\end{equation}
The relative angle between the field and the major axis of the
ellipsoid is obtained from
\begin{equation}\label{thetar}
\cos\theta_t=\cos\alpha\sin\theta\cos\phi+\sin\alpha\cos\theta.
\end{equation}
Since all azimuthal angles are possible $\left\langle \cos\phi
\right\rangle=0$, while $ \left\langle \cos^2\phi \right\rangle=1/2$;
and from eq.(\ref{ptheta}) $ \left\langle\cos^2\theta
\right\rangle=d^2/8b^2$. Thus, we finally arrive at
\begin{equation}\label{avecos}
\left\langle\cos^2\theta_t \right\rangle={d^2 \over 8b^2}+
{\cos^2\alpha \over 2}\left( 1-{3d^2 \over 16b^2} \right),
\end{equation}
which describes the MC anisotropy, when substituted in
eq.(\ref{eflux}).

\par\noindent{\bf (iii)} Another limit that is easily
accessible is for $d\simeq a$.  Now all the ellipsoids lie in the
plane, and $\cos\theta=0$, leading to $\left\langle\cos^2\theta_t
\right\rangle=\cos^2\alpha/2$. From eq.(\ref{eflux}) we then obtain
\begin{equation}\label{egeneral}
\langle \Phi^2\rangle={B^2\pi^2a^2\over 2}
\left[ 2 b^2-\left( b^2-a^2\right)\cos^2\alpha\right].
\end{equation}
This particular limit is plotted in Fig.\ \ref{fig9}, and can be
compared to experimental data for the angular dependence of MC (at
fixed temperature and field strength). The parameters $a=\sqrt{t\xi}$
and $2b=t$ in this figure are chosen to correspond with those reported
in ref.\onlinecite{rOvadyahu}.  The general form of the curve agrees
qualitatively with the experimental data. Another pertinent comment
concerns the experimental data of Laiko et al\cite{rbunchII}:
they compare results for fields parallel and perpendicular to the
current direction (while the field is in the plane of the sample).  On
averaging over hop directions, the MC for these configurations should
be identical, in agreement with experiments, except for very
disordered samples where $\ell$ is so large that averaging is
not appropriate.

\par\noindent{\bf (iv)} Finally for $d<a $, the NSS cigar
is flattened into a pancake and is no longer ellipsoidal. The
behavior is generally two dimensional, with $\Phi=\pi ab\times
B\sin\alpha$. This formula breaks down only at very small angles such
that $\alpha\leq d/b$, for which $
\left\langle \Phi^2 \right\rangle\approx (B\pi da)^2/2$.

Clearly, the general tendency is that as temperature is reduced (hence
$2b=t$ is increased), the ratio $d/t$ gets smaller. As indicated by
the above sequence, this leads to more and more pronounced effects in
the MC anisotropy. This is indeed consistent with the experimental
observations. In principle, the temperature dependence of this
increase could be measured experimentally and compared to the above
theoretical formulas.

\section{Discussion}  \label{Dis}
In this paper we studied the NSS model for quantum interference
effects of a tunneling electron in $D=3$. The three dimensional
geometry allows us to consider the relative orientations of the hop
and the magnetic field.  The effect of both spin-orbit active and
inactive impurities were taken into account. The results indicate
that, in the absence of SO impurities, there is positive MC due to an
increase in the localization length for fields both parallel and
perpendicular to the hop direction.  Furthermore, the MC data for
different fields collapses onto universal curves using the scaling
variables $B_{\perp}t^{3/2}$, and $B_{\parallel}t$ for the two
orientations.  This implies that, at least in the low field regime,
crossover effects are controlled by the flux penetrating an NSS
`cigar' whose transverse size grows diffusively (as $\sqrt{t}$).

In the presence of SO impurities, the numerical results again indicate
a positive MC, but as in two dimensions, no change in the
localization length is observed. No reduction in the fluctuations is
obtained in agreement with the replica arguments\cite{rMedinaRev}. The
results in this case are the same as those obtained from an
independent path approximation. The MC grows initially as $B^2$ for
both parallel and perpendicular orientations as expected.  Its
anisotropy disappears for large enough fields as the MC saturates to
roughly the same constant in both directions.  This could be checked
experimentally in the thin film limit ($d<t$) for samples doped with
heavy elements.

The most spectacular manifestation of these results is the possibility
of observing {\it bulk MC anisotropy}, when the sample is small enough
(or the characteristic length $\ell=\xi(T_0/T)^{(\nu+1)/(D+1)}$ is
large enough) that only a single hop (or just a few) dominates the
macroscopic conductance. The sample will then exhibit a `finger-print'
in its MC anisotropy, and the random orientation of the critical hop
can be determined by bulk anisotropy measurements. (It is possible
that this phenomenon explains why the experiments of Laiko et
al\cite{rbunchII} go from isotropic to anisotropic behavior as
disorder is increased\cite{rNota}.)  Large values for $\ell$ may be
achieved by either choosing samples with lower density of states at
the Fermi level, or larger $t/\xi$. Other manifestations of the {\it
disorder length scale} (including bulk MC anisotropy) may occur under
strong voltage bias\cite{rNguyenShklovskii}, if the electric field
reorients the critical hops; constraining the average over hop
orientations to within a cone.

However, the conductance of most samples with $L\gg\ell$ is governed
by many hops. Our estimate of $\ell$ based on published experimental
data\cite{rOvadyahu} indicates that it is much smaller than typical
sample planar dimensions.  Thus, averaging over many hop orientations
is inevitable, washing out the predicted anisotropy finger-print of a
single hop. Nevertheless, anisotropic behavior is still expected for
sufficiently thin samples (or at low temperatures). This is because
when the length of the cigar becomes comparable or larger than sample
thickness ($t>d$), the hops are forced to lie mostly parallel to the
sample plane. This restriction on hop orientations then leads to an MC
anisotropy that becomes more and more pronounced upon lowering
temperature. Appearance of such anisotropy in thin films has already
been observed in insulating $InO$ samples\cite{rOvadold,rJapones}. In
principle, the variations of anisotropy with temperature can be
measured accurately and compared to the formulas derived in the
previous section. Measurements of anisotropy can thus provide an
additional experimental tool for tests of quantum interference models.

\acknowledgements
We have benefited from discussions with R. F. Angulo and M. Araujo.
MK is supported by the NSF through grant number DMR-93-03667.  EM
thanks INTEVEP S.A. for permission to publish this paper. This work
was supported by the MIT-INTEVEP Collaborative Research Agreement.

\begin{figure}
\caption{The NSS model on a three dimensional diagonal
lattice. Impurities are located on the sites.}
\label{fig1}
\end{figure}

\begin{figure}
\caption{The figure depicts how effective areas that arise from scaling
can be derived from random walk arguments.}
\label{fig2}
\end{figure}

\begin{figure}
\caption{Log-conductance as a function of system size $t$ for fields
parallel to the hop direction. The change in the slope with the field
indicates an exponential correction to the conductance (change in the
localization length). A straight line is drawn as a guide to the eye.
The inset shows a reduction of fluctuations with the field. The power
law dependence on the hopping length is also indicated.}
\label{fig3}
\end{figure}

\begin{figure}
\caption{The figure shows the collapse of the log-conductance data
with the appropriate scaling variables (a) $Bt$ in the case the field
is parallel to the hop and (b) $Bt^{3/2}$ for fields perpendicular to
the hop direction. The scaling variable tells about the relevant area
threaded by the field.}
\label{fig4}
\end{figure}

\begin{figure}
\caption{Field dependencies for parallel and perpendicular field
directions when $Bt^{3/2}>1$ and for a single critical resistor.}
\label{fig5}
\end{figure}

\begin{figure}
\caption{Magnetoconductance in the presence of spin-orbit scattering.
The relevant log-conductance is no longer linear in $t$ for large
hopping lengths, indicative of no changes in the localization length
due to the field.}
\label{fig6}
\end{figure}

\begin{figure}
\caption{Collapse of the MC data in the presence of SO scattering. The
MC is governed by the areas a) $(\xi^{1/2}t^{1/2})^2$ for fields
parallel to a single critical hop and b) $\xi^{1/2}t^{3/2}$ for fields
perpendicular to the hop direction.}
\label{fig7}
\end{figure}

\begin{figure}
\caption{Angle convention used to perform averages over critical hop
directions. The magnetic field is along the $z$ axis making an angle
$\theta_t$ with the principal axis of the ellipsoid containing
dominant paths. The relevant magnetic flux penetrates through the
largest cross section of the ellipsoid perpendicular to the magnetic
field i.e. an ellipse of minor axis $a$ and major axis $c$.}
\label{fig8}
\end{figure}

\begin{figure}
\caption{Theoretical curve computed from eq.(18). The experimental data
correponds to $\xi = 85\AA$, $t=280\AA$ and sample thickness $d=250\AA$.
Notice how the curve fall towards $\alpha = 0$ (parallel to sample),
closely resembling the experimental data of reference [2]. }
\label{fig9}
\end{figure}

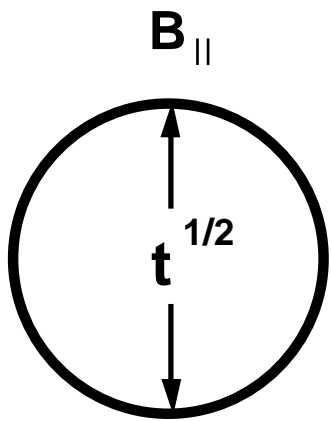
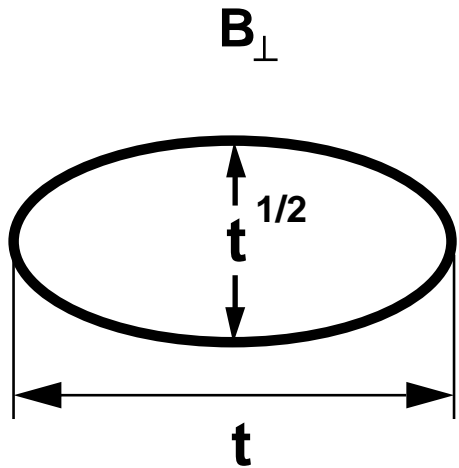
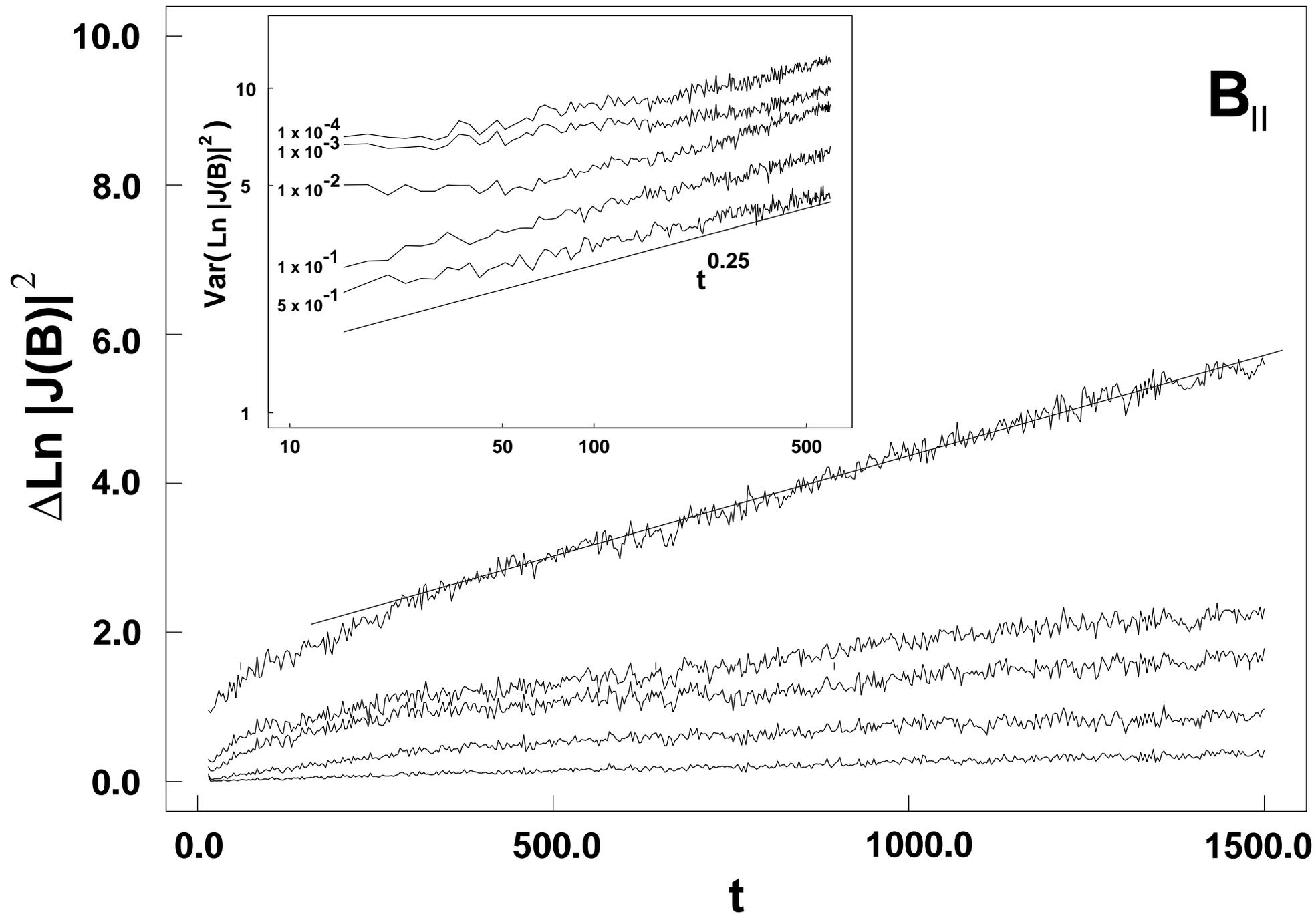
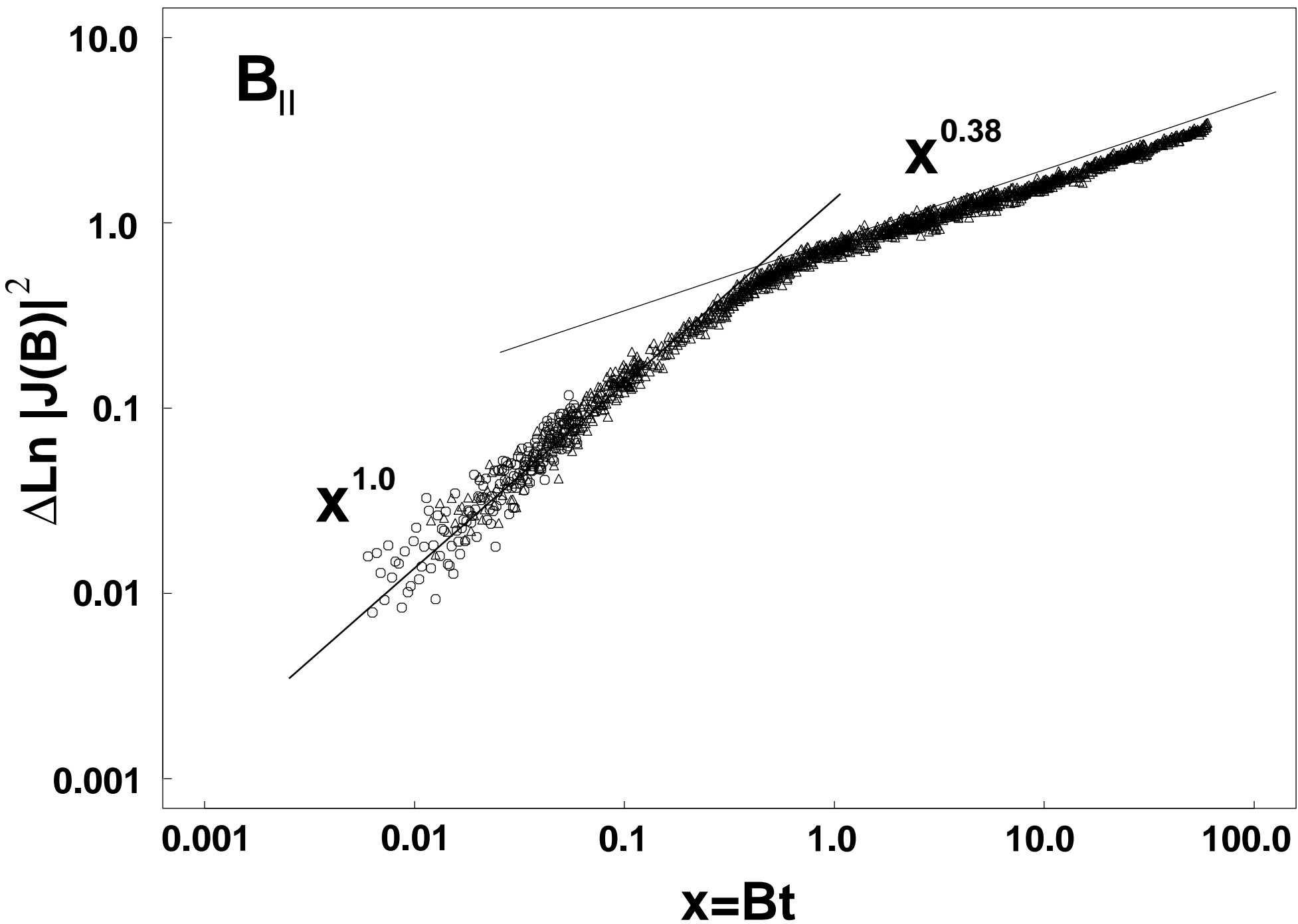
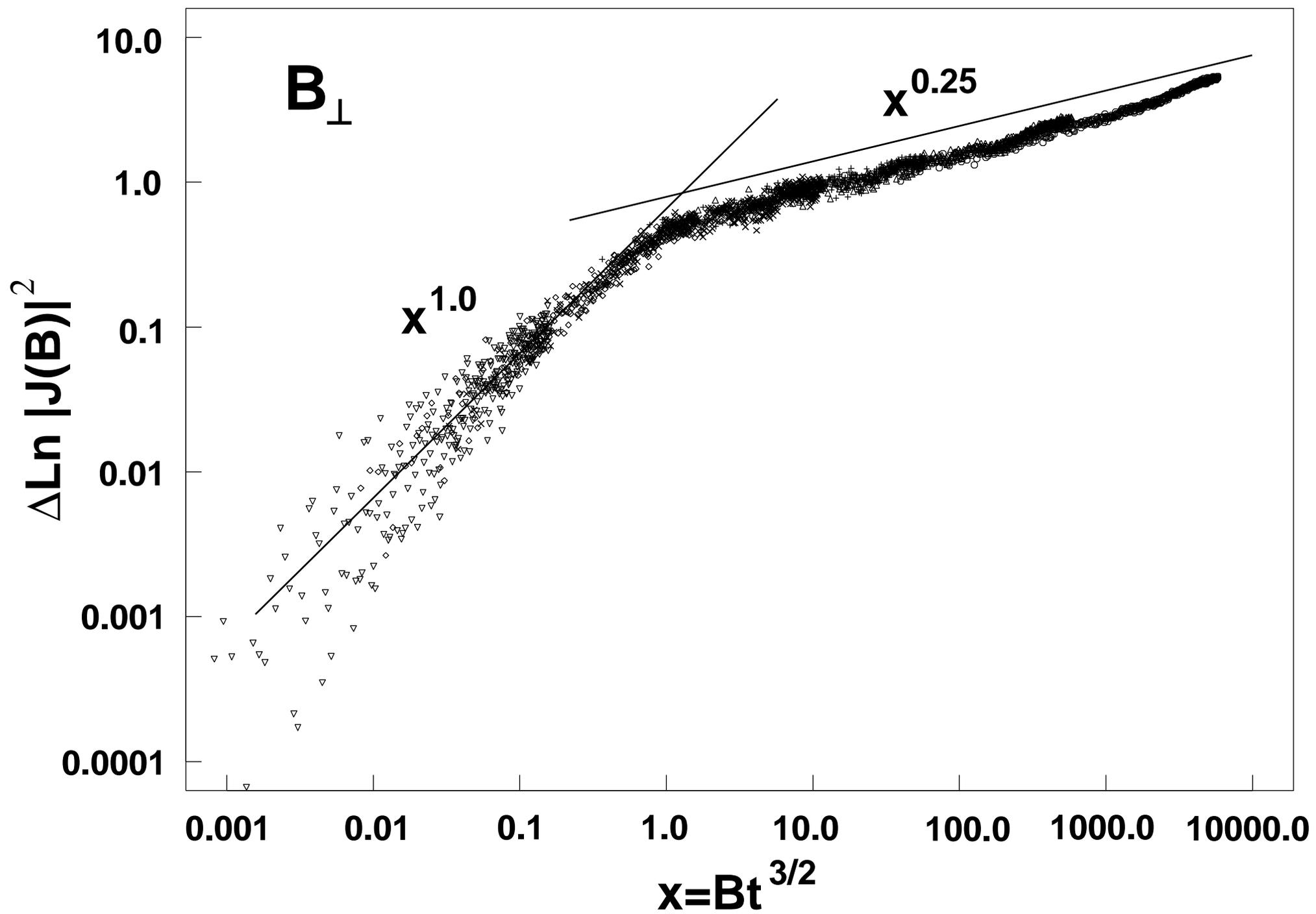
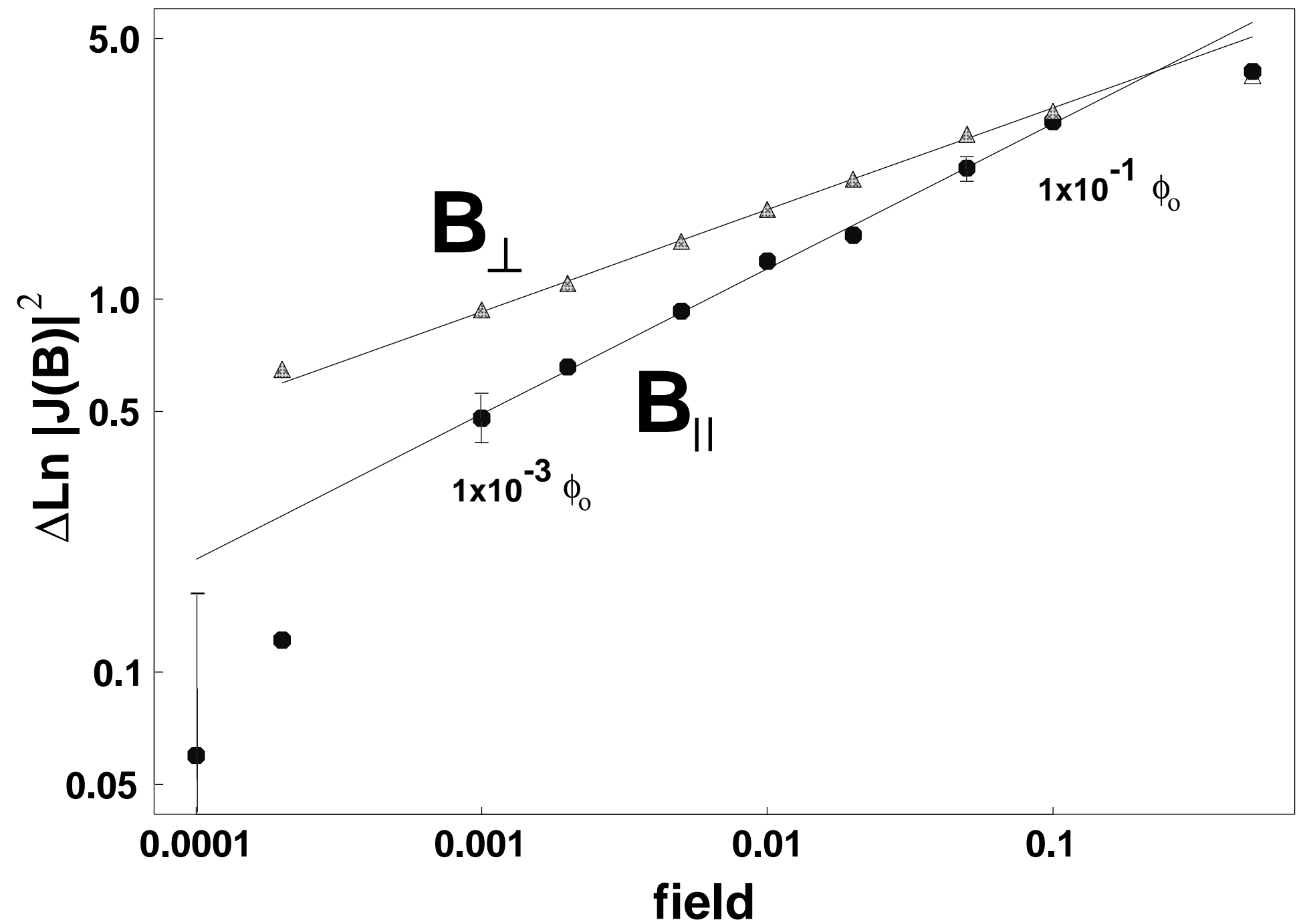
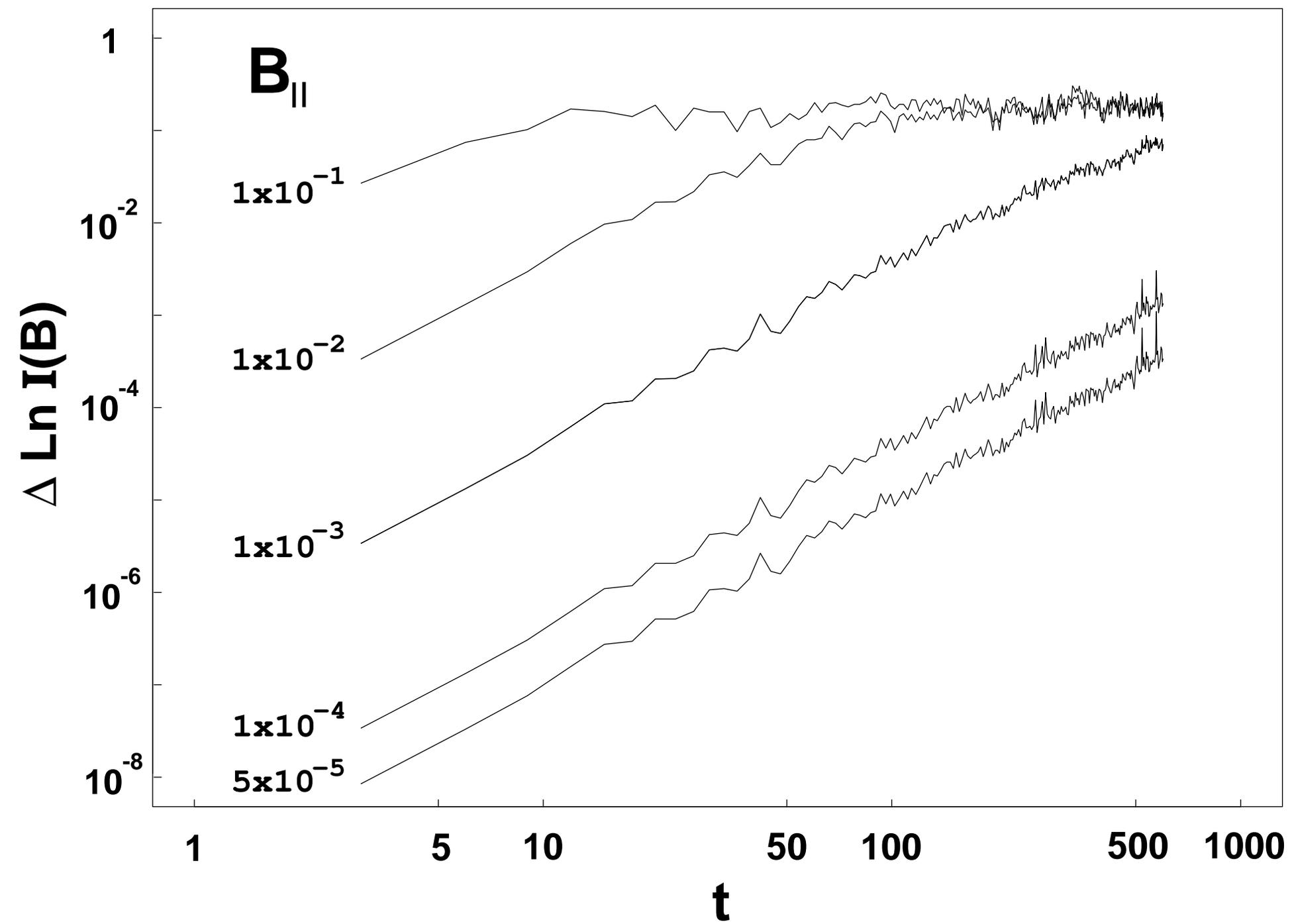
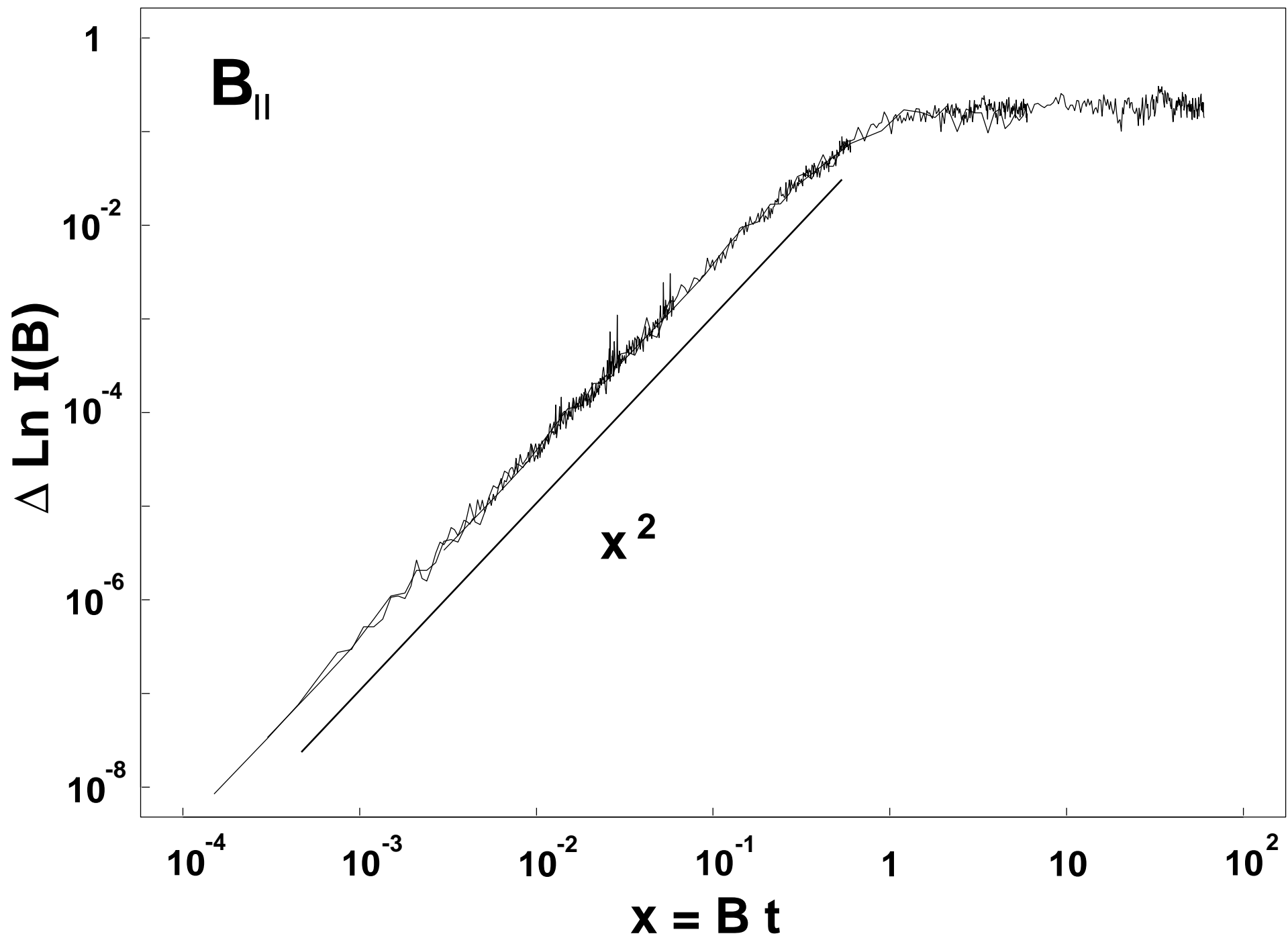
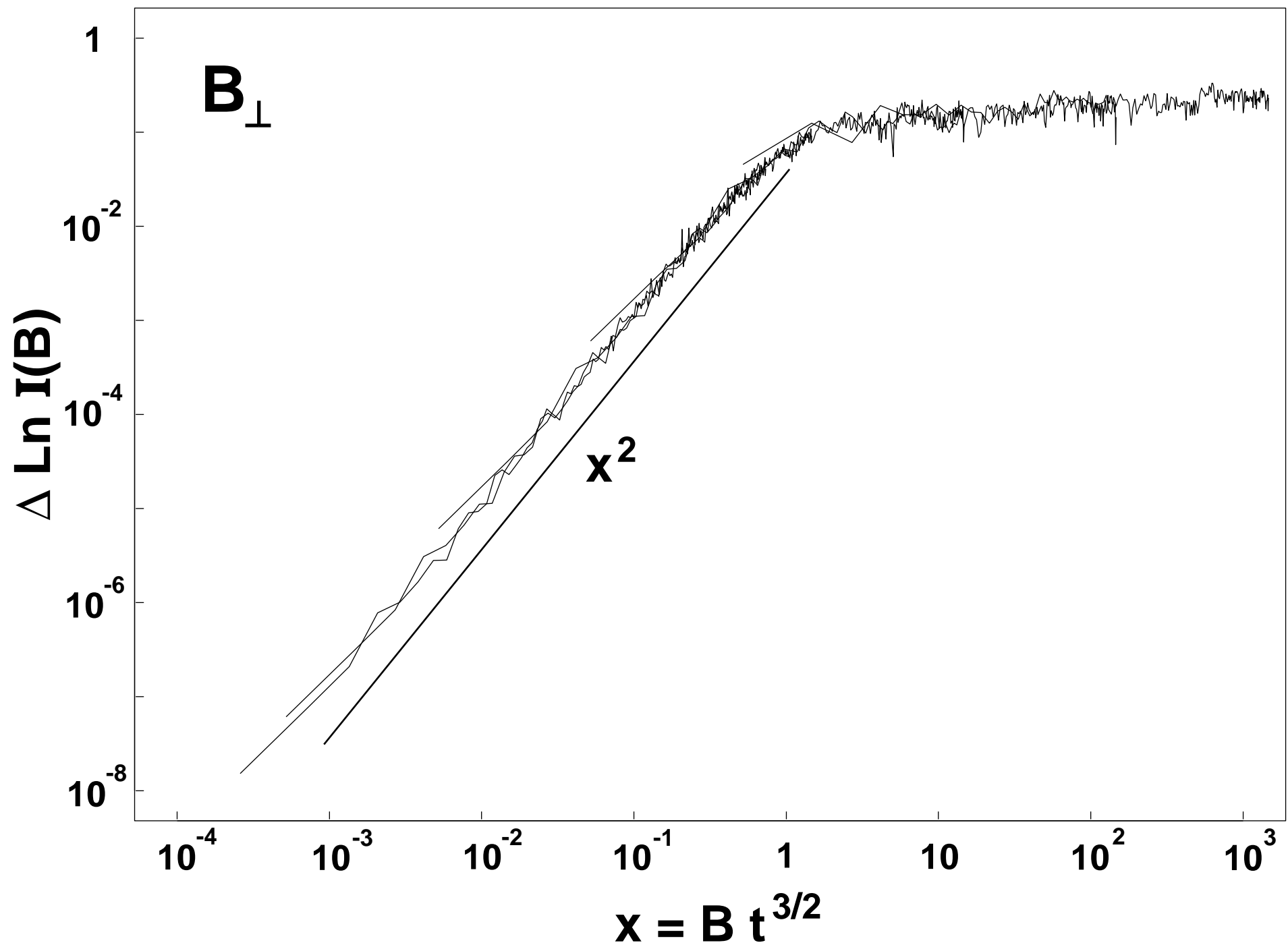
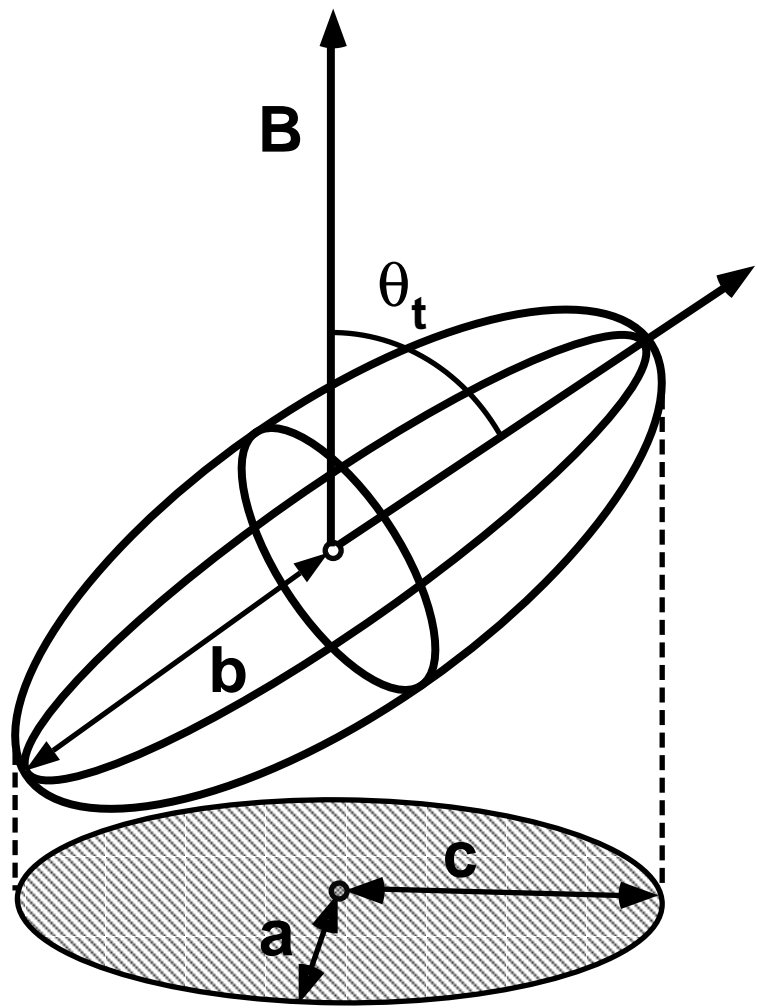
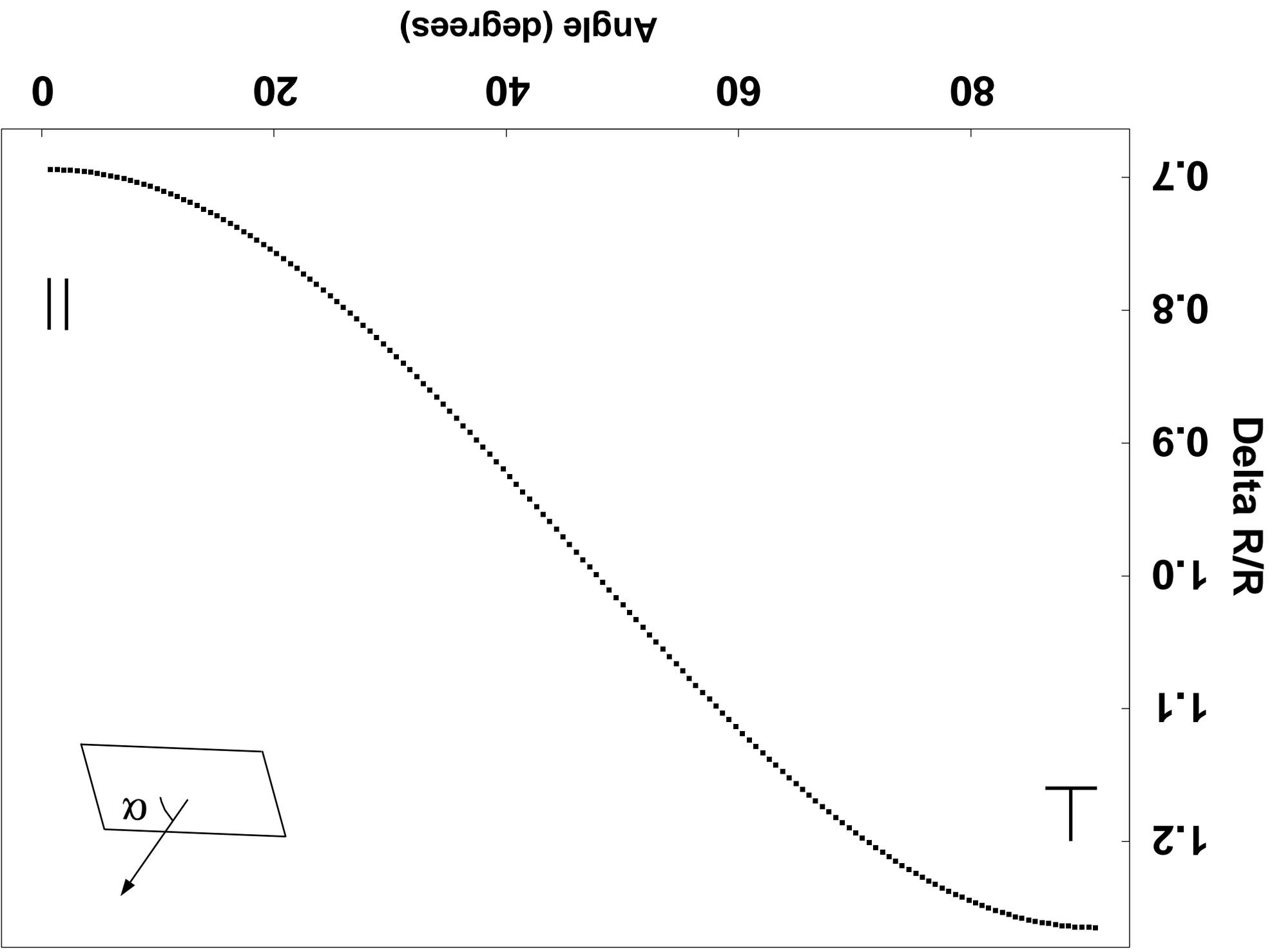
\end{document}